\documentclass[twocolumn,showpacs,preprintnumbers,amsmath,amssymb,floatfix]{revtex4}
\usepackage[dvips]{graphics}
\usepackage{ lscape, longtable, bm}
\begin{document}
\title{Mediation of Long-Range Attraction Selectively between Negatively-Charged Colloids on Surfaces by Solvation}
%Interfacial Coexistence of Dilute Phases and Periodic Structures in Charged Colloidal Suspensions}
\author{William Kung}
\email{w-kung@northwestern.edu}
\affiliation{Department of Materials Science and Engineering, Northwestern University, Evanston, Illinois 60208-3108, USA}
\author{Monica Olvera de la Cruz}
\email{m-olvera@northwestern.edu}
\affiliation{Department of Materials Science and Engineering, Northwestern University, Evanston, Illinois 60208-3108, USA}

\begin{abstract}

We propose a mean-field analytical model to account for the observed asymmetry in the ability to form long-range attraction by the  negatively charged colloidal particles and not their equivalently charged positive counterpart.  We conjecture that this asymmetry is due to solvation effects, and we phenomenologically capture its physics by considering the relative strength of this water-induced short-range repulsion between the different charge species.  We then apply our model to the colloidal system of negatively charged disks that are neutralized by a sea of counterions and strongly absorbed to an interface in a compressible binary system.  We demonstrate the resulting coexistence between a dilute isotropic ionic phase and a condensed hexagonal lattice phase as a function of density and interaction strength.  
\end{abstract}

\date{\today} 

\maketitle

\def\coth{\hbox{\rm{coth}}}
\def\csch{\hbox{\rm{csch}}}
\def\sech{\hbox{\rm{sech}}}

Colloidal particles have always been the prime constituents of prototypical self-assembling systems.  Their phase properties can be easily probed by optical techniques and readily manipulated via chemical means in the laboratory.  Their rich chemistry leads to such diverse applications as uses in coating materials, ceramic precursors, and the designing and manufacturing of biological macromolecules.  It has been found that, biologically, absorption of colloids into liposomes can be used to stabilize vesicles~\cite{Granick2007}.  The collective behavior of colloidal particles has also been recently demonstrated to provide a coorperative probe of membrane surfaces~\cite{Groves}.  

Specific colloidal interactions that influence the system's overall properties include Van der Waals dispersion forces, which are the direct consequence of the induced dipole-dipole interactions created from the quantum fluctuation in the charge densities~\cite{Parsegian2006}, electrostatic interactions between the charge densities themselves~\cite{Verweytext}, as well as an effective attraction of the larger particles due to depletion when the colloidal suspension contains particles of different sizes~\cite{Russeltext}.  At the mean-field level where we neglect fluctuation effects, it is well known that the Poisson-Boltzmann framework precludes long-range attraction between like-charges, in contradiction to experimental findings~\cite{PBviolations} such as those found in confined systems~\cite{Grier2003}.  So one may account for these long-range attraction between like charges by going beyond mean-field approaches and take in account fluctuation effects.  

However, there has been recent experimental indications that this long-range attraction occurs only between colloidal particles of negative charge and not of the opposite kind~\cite{Groves}.  Depletion cannot account for this asymmetry since it is entropic in nature and should not distinguish between the two kinds of charge.  On the other hand, Van der Waals forces are much smaller in magnitude by comparison to be a possible significant contributing factor. 

In this paper, we propose a simple, mean-field analytical model that would demonstrate this asymmetric behavior between the negatively (positively) charged colloidal disks in their ability (inability) to form long-range attraction based on their interaction with water molecules.  In particular, we conjecture that due to the bent-core shape and the charge distribution of the water molecule, the greatest aggregation of them -- {\it{swelling}} -- occurs between the larger-sized colloidal particles that are negatively charged and the smaller-sized counterions that are positively charged.  Swelling also occurs, albeit to a lesser degree, in the opposite case where the system consists of equivalently charged positive particles with negative counterions.  In contrast, swelling would occur much less favorably between particles that are of the same charge type no matter their respective size.  Therefore, this interaction with water molecules induces an effective {\it{repulsion}} between {\it{opposite}} charges and correspondingly a comparative {\it{attraction}} between {\it{like}} charges, which exactly counter-balances the effects of electrostatics.  Consequently, there exists a general competition between these two types of interaction in the system.  For a range of interaction parameters and densities, we will demonstrate the resulting coexistence of a low-density isotropic phase and a high-density hexagonal lattice phase in our mean-field model.

\begin{figure}[t]
\vspace{2mm}
\hspace{23mm}
\includegraphics{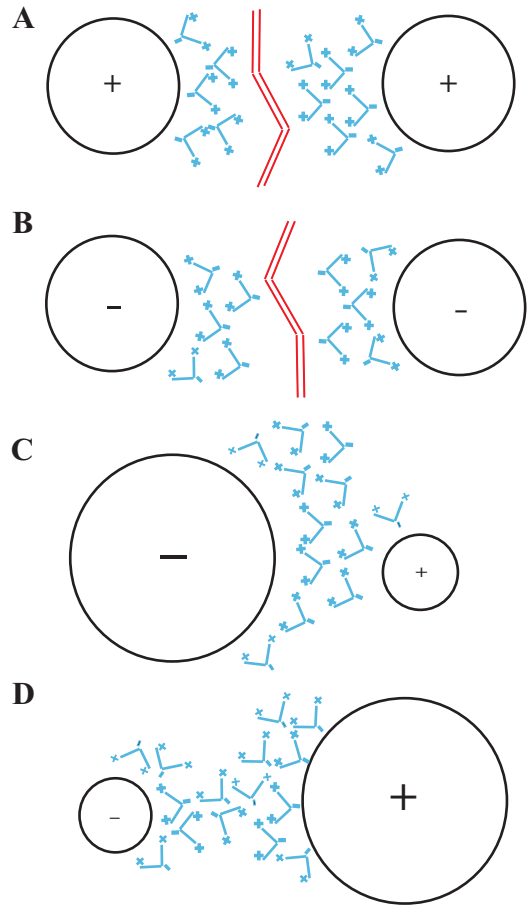}
\vspace{2mm}
\caption{Interactions between the bent-core water molecules (blue) and the positive and negative particles.  There exists great frustration in the arrangements of water molecules around neighboring like-charge particles in (A) and (B). Due to the bent-core geometry of the water molecules, they arrange themselves in more favorable configurations of the larger negative particles with smaller positively charged particles (C) than of the reverse scenario (D).}
\label{Solvation}
\end{figure}

As mentioned, the interaction between the colloidal particles and water molecules is central to our conjecture in accounting for the observed attraction between negatively  charged colloidal particles and the lack thereof between positively charged ones.  To understand this asymmetry, we consider the four scenarios represented in Fig.~\ref{Solvation}.  In the first two situations respectively depicted in Figs.~\ref{Solvation}A and~\ref{Solvation}B when we have like-charge particles in solution, there exists great frustration in the arrangement of water molecules around neighboring particles irrespective of the sign on the charge of these particles.  While it is much more preferable for water molecules to situate themselves between oppositely charged particles, there does exist a difference in the two cases in which the larger-size particles are negatively charged particles (Fig.~\ref{Solvation}C) and in which they are positively charged (Fig.~\ref{Solvation}D).  Due to the bent-core geometry of the water molecules and the charge distribution on the molecules (positive charge-concentration on the hydrogen atoms and negative charge-concentration on the oxygen atom), it would be easier for water molecules to orient and anchor themselves more energetically favorably between a larger particle that is negatively charged and the corresponding counterion(Fig.~\ref{Solvation}C) than in the reverse scenario when the larger particle is positively charged (Fig.~\ref{Solvation}D).  Thus, we would expect the strongest swelling in the scenario depicted in  Fig.~\ref{Solvation}C and a correspondingly weaker case in the scenario depicted in Fig.~\ref{Solvation}D.  If we parametrize this water-induced short-range repulsion between the charged particles by $w_{ij}$, for colloidal particles of type $i=\{\pm\}$ and counterions of type $j=\{\pm\}$, we would expect our conjecture to predict the general relation $\{w_{++}, w_{--}\}<w_{+-}<w_{-+}$.  The detailed elucidation of their magnitude would necessitate finer knowledge in such microscopic details regarding the interaction between the colloidal particles and water molecules as their precise charge distributions, their associated (quantum) fluctuations, and the molecular geometry of the bent-core water molecules.  Our conjecture is in line with the theoretical calculation presented in~\cite{Pedro2006} for the effective pair potentials (EPPs) between ions immersed in water based on the reference interaction site model (RISM) approach~\cite{Pedro2002}.  There the EPPs were found after contracting out the degrees of freedom associated with the oxygen and hydrogen sites within the water molecules.  In their system of NaCl, Gonzalez-Mozuelos found that $w_{++}<w_{--}<w_{-+}$.  Though in his case, the choride ions are large than the sodium ions by default, and the quantity $w_{+-}$ was not applicable.   

In what follows, we will present a phenomenological minimal model that would capture the selective formation of long-range attraction by the negatively charged particles in aqueous solution.  We will demonstrate that our model does indeed yield coexistence between a dilute phase and a condensed phase when the ratio of the water-induced repulsion strengths, $w_{\pm\mp}/w_{\pm\pm}$, between opposite and like charges becomes high enough.  For concreteness, we will consider a compressible  binary system consisting of charged colloidal disks of radius $R$ submerged in aqueous solution of counterions of radius $a$.  Based on our conjecture, when the colloidal disks are negatively charged, we would expect the magnitude of $w_{-+}\,(R_->a_+)$ to be larger than $w_{+-}\, (R_+>a_-)$, which corresponds to the opposite case in which the colloidal disks are positively charged.  Again, our model cannot predict their magnitudes without incorporating further the molecular details regarding the particles' interaction with water.  Without loss of generality, we will now focus on the case when the larger colloidal particles are negatively charged and demonstrate their formation of an ordered phase in our model.

For the dilute isotropic phase, its homogeneity and low density of charged particles permit its treatment as a two-dimensional charged gas in a homogeneous background that displays nonselective interactions with the charged components.  As previously mentioned, we have two competing interactions in the system:  the water-induced short-range repulsion between the charged disks and the electrostatic interaction between them.  Based on the EPPs numerically computed in the RISM framework~\cite{Pedro2006, Pedro2002}, we analytically model the short-range repulsion between the counterions and colloidal disks as a Gaussian curve enveloping the numerically plotted functional forms of their EPPs~\cite{Pedro2006, Pedro2002}, which in Fourier space would take the form
\begin{eqnarray}
U^{sr}_{\bm{k}}&=&4\pi R^2w_{--}\,e^{-k^2R^2}+4\pi a^2w_{++}\,e^{-k^2R^2}\nonumber\\
&&+8\pi aR\,w_{-+}\,e^{-k^2aR}\;,
\label{SR}
\end{eqnarray}
in units of $k_BT$, The electrostatic interaction between them, on the other hand, is simply
\begin{eqnarray}
U^{el}_{\bm{k}}&=&\frac{Z_iZ_j\ell_B}{k}\;;\,\,\,\,\,\,\,\,\,i,j = +,-,
\label{EL}
\end{eqnarray}
where $\ell_B$ is the Bjerrum length and takes the value of $7\AA$ in the systems henceforth considered.  Here we will simply approximate the electrostatic interaction between the smeared-out charge distribution contained in the colloidal disks as that between their centers of mass and thus assume a Coulomb form of interaction.  Further, the configuration of the system is such that we can assume the three-dimensional form for this interaction.  Refinement to the above assumption can be straightforwardly performed for other systems, such as those with salt.  

Since we only consider the mean-field limit in our model, the effects of electrostatics, which only comes in through the one-loop order, would not contribute to the gas phase.  Therefore, the gas-phase free energy, $f^G$,  has two contributions:  water-induced short-range repulsion and entropy.  It takes the form
\begin{eqnarray}
\beta f^G\vert_{\bm{k}=0}&=&\rho_1\log\frac{\rho_-}{e}+\rho_+\log\frac{\rho_+}{e}+\frac{\rho_-^2}{2}\left(4\pi R^2\right)w_{--}\nonumber\\
&&+\frac{\rho_+^2}{2}\left(4\pi a^2\right)w_{++}+\rho_+\rho_-\left(4\pi a R\right)w_{-+},
\label{GasFreeEnergy}
\end{eqnarray}
where the colloidal number density $\rho_-$ and counterion density $\rho_+$ satisfy the electro-neutrality condition, $Z_+\rho_++Z_-\rho_-=0$.  Again, the interaction between the water molecules and the colloidal particles can be encoded directly into the parameters $w_{ij}$ characterizing the strength of these water-induced short-range repulsive interactions between the colloidal disks and counterions. Taking into account our conjecture, we simply assume that $w_{++}=w_{--}<w_{-+}$.  

\begin{figure*}[t]
\vspace{2mm}
\hspace{5mm}
\includegraphics{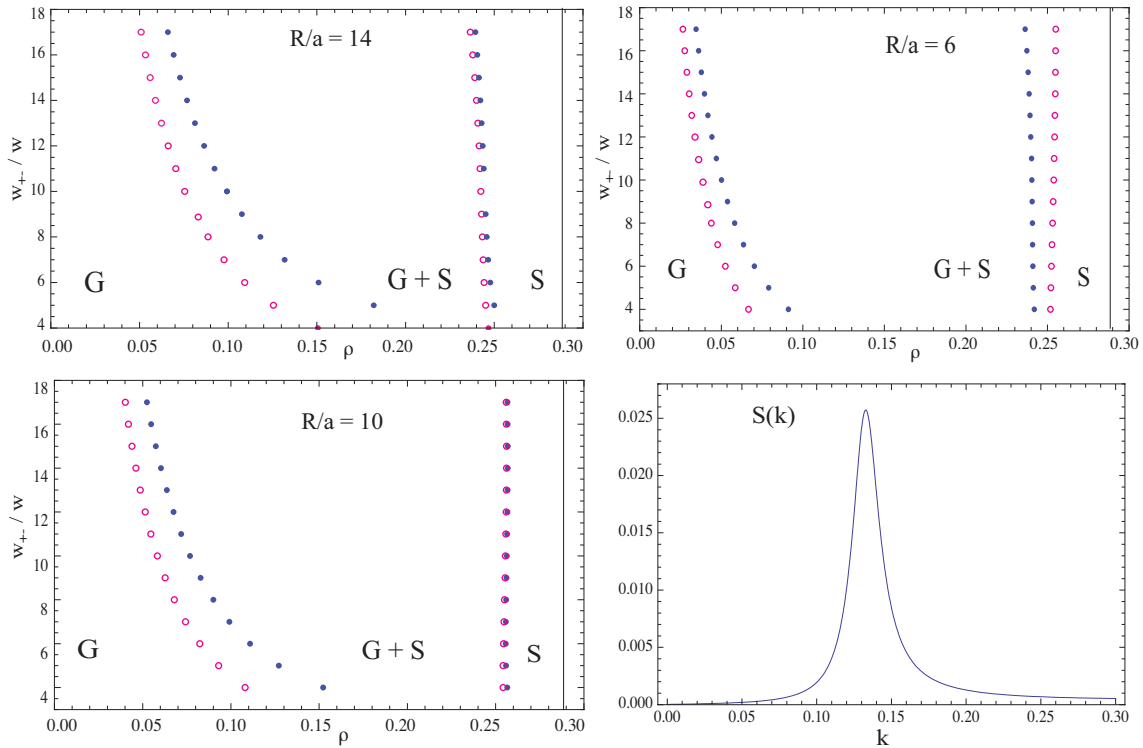}
\vspace{2mm}
\caption{Phase Diagrams and Structure Factor.  We show here the phase diagrams for size ratios of $R/a =14$, $R/a=10$, and $R/a=6$ and for charge ratios of $\vert Z_1/Z_2\vert=15$ (blue dots) and $\vert Z_1/Z_2\vert=10$ (red circles).  Starting from low colloidal number density $\rho$, we progress from the gas phase (G), to the coexistence region (G+S), to the solid phase (S).  The close-packing number-density is $\rho_c=0.2886/R^2$ (grey vertical line), which corresponds to the two-dimensional hexagonal close-packing volume fraction of $\varphi=0.9069$.  Here we assume that $w_{++}=w_{--}=w$.  The structure factor is related to the effective interaction potential via $S^{-1}(\bm{k})=U_{--}(\bm{k})+\rho_-^{-1}$ in the low-density limit.  This is done for a sample of colloidal disks with a size ratio of $R/a=14$, a charge ratio of $\vert Z_-/Z_+\vert=10$, $w_{-+}/w=6.8$ and at a number density of $\rho=0.127/R^2$ within the coexistence region. }
\label{PhaseDiagram}
\end{figure*}

To demonstrate long-range attraction between negatively charged colloidal disks, we integrate out the fast modes associated with the smaller-sized counterions in order to obtain the effective interaction potential between the colloidal disks~\cite{Olvera1994}
%
%\begin{widetext}
\begin{eqnarray}
U_{--}(\bm{k})&=&\frac{\pi Z_-^2\ell_B}{k}+4\pi R^2\,w_{--}\,e^{-k^2R^2}\nonumber\\
&&-\left[\frac{\left(\frac{\pi Z_-Z_+\ell_B}{k}-4\pi aR\,w_{+-}\,e^{-k^2aR}\right)^2}{\frac{\pi Z_+^2\ell_B}{k}+4\pi a^2\,w_{++}\,e^{-k^2a^2}+\frac{1}{\bar{\rho}_+}}\right]\;.
\label{A}
\end{eqnarray}
%\end{widetext}
%
The structure factor $S(\bm{k})$, in turn, is related to the effective interaction potential in Eq.~\ref{A} via $S^{-1}(\bm{k})=U_{--}(\bm{k})+\rho_-^{-1}$.  For a range of parameters, we do indeed observe the corresponding peak in the structure factor for this system.  An example of which is shown in Fig.~\ref{PhaseDiagram}.  Thus, we do see correlation over certain length scale between the negatively charged colloidal disks in the gas phase of our system, a prelude to actual crystallization.  
%
%\begin{figure}[t]
%\vspace{2mm}
%\hspace{23mm}
%\includegraphics{structurefactorB}
%\vspace{2mm}
%\caption{Momentum dependence of the structure factor $S(\bm{k})$.  The structure factor is related to the effective interaction potential via $S^{-1}(\bm{k})=U_{--}(\bm{k})+\rho_-^{-1}$ in the low-density limit.  This is done for a sample of colloidal disks with a size ratio of $R/a=14$, a charge ratio of $\vert Z_-/Z_+\vert=10$, $\epsilon_{-+}/\epsilon=6.8$ and at a number density of $\rho=0.127/R^2$ within the coexistence region.}
%\label{structurefactor}
%\end{figure}
%
%

For the periodic phase, it has been established that the optimal structure in the case of asymmetric charge ratios between the colloidal particles and counterions in two dimensions is the hexagonal closed-packed lattice~\cite{Loverde2007}.  We model the two-dimensional closed-packed hexagonal lattice using a foam model developed earlier to study non-closed packed lattices in colloidal crystals~\cite{Kung2002, Kung2004}.    In the foam model, the free energy for the solid is divided into an entropic contribution and an interaction contribution.  To compute the entropic contribution, we follow the method developed earlier~\cite{Ziherl2000, Ziherl2001} by adopting the cellular free-volume theory.  In this approximation, each colloidal particle is confined to a cage formed by its neighbors.  The free volume available to each particle's center of mass is the volume of the Wigner-Seitz cell after a layer of thickness of $\sigma$ (where $\sigma$ is the hard-core radius of particles) is peeled off its faces.  Using simple geometric arguments, the cellular free-volume approximation yields the following entropy contribution to the free energy in the high-density limit:
\begin{eqnarray}
\beta f_e&=&-\log\left(\frac{1}{2\sqrt{3}\rho_-}-1\right)\;.
\end{eqnarray}
Despite its mean-field nature, the free volume theory yields excellent quantitative agreement with available numerical simulations in the high-density limit~\cite{Ziherl2000, Ziherl2001}.  

On the other hand, the interaction contribution to the solid free energy can be carried out by direct summation of the reciprocal-lattice series~\cite{Olvera2005}. To compute the interaction energy per unit volume between the colloidal disks in the hexagonal lattice, we use the effective colloidal potential in Eq.~\ref{A} and apply the Poisson summation formula that replaces the sum over the lattice vectors $\bm{\Lambda}$ by a sum over the reciprocal lattice vectors $\bm{Q}$ defined by the property $\bm{Q\cdot \Lambda}=2\pi m$ for some integer $m$.  We thus obtain
\begin{eqnarray}
\beta f_{int}&=&\frac{1}{2\mathcal{A}^2}\sum_{\bm{Q}}\hat{\sigma}\left(\bm{Q}\right)^2U_{--}\left(-\bm{Q}\right)\;,
\end{eqnarray}
where $\mathcal{A}$ is the area of a unit cell and $\hat{\sigma}(\bm{k})$ represents the transform of a circular-domain charge density
\begin{eqnarray}
\hat{\sigma}(\bm{k})&=&\frac{2\pi R\rho_-}{k}J_1(kR)\;,
\end{eqnarray}
and $J_1(.)$ is the regular Bessel function of order $1$.  In total, the solid free energy is simply $\beta f^S\vert_{\bm{k}=0}=\beta f_e+\beta f_{int}$.  Together with the gas free energy $\beta f^G\vert_{\bm{k}=0}$ in Eq.~\ref{GasFreeEnergy}, we can determine phase coexistence of the periodic solid and dilute gas by using the common-tangent rule by imposing equal chemical potential and pressure for the two phases.  The resulting phase diagrams are shown in Fig.~\ref{PhaseDiagram} for particle size-ratios of $R/a = \{6, 10, 14\}$ and charge-ratios of $\vert Z_-/Z_+\vert = \{10, 15\}$. For all probed size-ratios,  starting at low colloidal number density $\rho_-$  we progress from the gas phase (G), to the coexistence region (G+S), to finally the solid phase (S) at high density.  The close-packing number-density is $\rho_c=0.2886/R^2$, which corresponds to the two-dimensional hexagonal close-packed volume fraction of $f=0.9069\, (f=\pi R^2\rho)$.  Here we assume that $w_{++}=w_{--}=w$, since our conjecture makes no {\it{a priori}} distinction between the two cases regarding their effects in water.

From Fig.~\ref{PhaseDiagram}, we see that the coexistence region decreases as the charge ratio increases from $\vert Z_-/Z_+\vert =10$ to $\vert Z_-/Z_+\vert = 15$.  This is expected since when we increase the strength of the electrostatic attraction (Eq. \ref{EL}) between the negatively charged colloidal disks and the counterions, the water-induced attraction between the colloidal disks correspondingly decreases in relative strength, which allows for the increased gas region in the phase diagram.  We also see that the same effect when we keep the electrostatic interaction constant and decrease the size-ratios between the colloidal disks and counterions.  To understand this, we recall that the short-range repulsion we model in Eq. (\ref{SR}) depends explicitly on the sizes of the particles.  Therefore, decreasing their size-ratio is equivalent to decreasing the relative strength of their water-induced short-range repulsion.  Hence, the same reason that swells the gas phase region in the phase diagram with increasing charge-ratio also applies here as we decrease the size-ratio.  In this respect, our model provides concrete predictions that can be experimentally tested in its regime of applicability.  

Given the mean-field assumption in the construction of our gas and solid free-energy densities, our model does not indicate the existence of a critical point for the gas-solid transition in this compressible binary system involving colloidal disks and counterions.  At low enough ratios of $w_{-+}/w$ (as a function of size- and charge-ratios), we see that the gas phase would abruptly preempt the solid phase by having a globally minimum free energy for all densities.  The reverse is true in the opposite limit of high enough ratios of $w_{-+}/w$ where the solid phase would preempt the gas phase.  However, given that cellular free-volume approximation only works well in the high-density regime, we expect our model would no longer be valid in the phase-space region wherein the solid free energy starts dominating the gas free energy as the global minimum for all densities.  For example, in the case of size-ratio of $R/a=14$, the gas free energy becomes the global minimum for $w_{-+}/w < 5$, and our model indicates that the system would be completely gaseous.  For the smaller charge-ratio of $\vert Z_-/Z_+\vert=5$, we find that our model becomes unreliable when $w_{-+}/w>10$ for the size-ratio $R/a=6$ at which point the solid free energy dominates the gas free energy for all densities, even in the low-density regime where the free-volume approximation is less reliable.  Likewise, we approach the limit of applicability for our model when $w_{-+}/w>13$ for $R/a=10$, and $w_{-+}/w>15$ for $R/a=14$.

As discussed before, our model makes no  {\it{a priori}} predictions on the magnitudes of the water-induced short-range repulsion between different charge species.  We can only infer the ordering of their magnitude.  One possible refinement of our mean-field model is to generalize the short-range repulsion parameters, $w_{\pm\pm}$ and $w_{\pm\mp}$, as functions of particle size and molecular geometry.  We can also account for the self-interaction between water molecules.  Recently, there has been wide recognition in the important role water plays in many pertinent biological processes~\cite{water1}.  As a result, there has been much concerted effort on devising more accurate models for the various observed properties of water~\cite{water2}.

We note that short-range attractions between macroions in the presence of high-valence counterions also lead to phase segregation~\cite{LevinReview}.  These precipitation has been explained by various modified Poisson-Boltzmann approaches long ago~\cite{Stevens1990, Olvera1995, Liu1999, Levin1999, Olvera1999}.  Our present model of segregation is very different in nature since it involves long-range attractions among negatively charged colloids due to assymetric hydration forces.  Further, we have neglected charge renormalization here since ion condensation along the colloids has been found in dilute three-dimensional colloidal suspensions~\cite{Chaikin1984, Stevens1996};  in our system, the lattice phase is concentrated.  Moreover, charge renormalization would yield only negligible corrections to the theory.

It is our hope that our simple model provides a concrete example in which the molecular structure and chemical details of water contribute crucially to the understanding of the phase properties of such systems as the self-assembly of colloidal disks.  Having demonstrated another example of the inadequacy in treating water as a continuous, inert dielectric medium, it is our further hope that our model, with additional refinements, would serve as a starting point in modeling other more complicated, self-assembling systems such as charged colloidal suspensions in three dimensions~\cite{Kung2007}.

\begin{acknowledgments}
W. K. gratefully achknowledges stimulating conversations with Y. Velichko, D. Zhang, S. Loverde, P. Gonzalez-Mozuelos, and A. G. Roche.  This work was supported by the ACS PCF Grant 44645-AC7.
\end{acknowledgments}

\end{document}